\documentclass[12pt,a4paper]{article}
\usepackage[numbers,sort&compress]{natbib}
\usepackage{amssymb,amsmath,graphicx}

\newcommand{\vc}[1]{\mathbf{#1}}
\newcommand{\uvc}[1]{\mathbf{\hat #1}}

\newcommand{\dd}{\mathrm{d}}

\newcommand{\cn}{\mathop{\rm cn}\nolimits}
\newcommand{\dn}{\mathop{\rm dn}\nolimits}

\begin{document}
\title{\bfseries Twist of cholesteric liquid crystal cell
with substrates of different anchoring strengths}
\author{A.D.~Kiselev\\
Chernigov State Technological University\\
Shevchenko Str. 95,
14027 Chernigov, Ukraine\\
Email: kisel@elit.chernigov.ua
}
\date{}

\maketitle

\begin{abstract}
  We consider director configurations of cholesteric liquid crystal
  (CLC) cells with two plane confining substrates.  Exact solutions of
  the Euler-Lagrange equations for out-of-plane orientations of the
  easy axes that correspond to inhomogeneous conical structures of CLC
  director are derived.  We study dependence of the CLC twist
  wavenumber on the free twisting number assuming that anchoring
  energies at the substrates are either equal or different.  In both
  cases this dependence is found to be generally discontinuous with
  hysteresis loops and bistability effects involved.  For CLC cells
  with identical substrates and planar anchoring conditions the
  jump-like behaviour only disappears in the weak anchoring limit.
  Contrastingly, when the anchoring strengths are different, there is
  the finite value of anchoring below which the dependence becomes
  continuous.  Another effect is the appearance of the gap between the
  adjacent twist wavenumber intervals representing locally stable
  director configurations.  We calculate the critical value of
  anchoring asymmetry and present the results of numerical
  calculations.
\end{abstract}



\section{Introduction}
\label{sec:intro}

In equilibrium cholesteric phase molecules of a liquid crystal (LC)
align on average along a local unit director $\uvc{n}(\vc{r})$ that
rotates in a helical fashion about a uniform twist
axis~\cite{Steph:1974,Gennes:bk:1993}.  This tendency of cholesteric
liquid crystals (CLC) to form the helical twisting pattern 
is caused by the presence of anisotropic molecules with
no mirror plane~---~the so-called chiral molecules
(see~\cite{Harr:rmp:1999} for a recent review).

The phenomenology of CLCs can be explained in terms of
the Frank free energy
\begin{align}
&
F_b=\frac{1}{2}\int \dd^3 x \bigl\{
K_1 ({\nabla}\cdot\uvc{n})^2+K_2 (\uvc{n}\cdot\nabla\times\uvc{n})^2
\notag\\
&
+K_3\,[\uvc{n}\times(\nabla\times\uvc{n})]^2+ 
2h\,\uvc{n}\cdot\nabla\times\uvc{n}
\bigr\}\,,
\label{eq:frank}  
\end{align}
where $K_1$, $K_2$ and $K_3$ are the splay, twist and bend
Frank elastic constants. 
As an immediate consequence of the broken mirror symmetry,
the expression for the bulk free energy~\eqref{eq:frank}
contains the chiral term proportional 
to the chiral strength parameter $h$.
This parameter gives the equilibrium value of CLC twist wavenumber, 
$q_0=h/K_2$, which corresponds to the pitch
$P_0\equiv2\pi/q_0$. 
In what follows the parameter $q_0$ will be referred to as 
the free twist wavenumber or the free twisting number.
So, if the twist axis coincides with the
$z$-direction, the director field
$\uvc{n}=(\cos q_0 z, \sin q_0 z, 0)$ defines the equilibrium configuration.
Since $\uvc{n}$ and $-\uvc{n}$ are equivalent in liquid crystals,
periodicity of the spiral is given by the half-pitch, $P_0/2$.
 
Typically, the pitch $P_0$ can vary from hundreds of nanometers to
many microns or more, depending on the system.  The macroscopic chiral
parameter $h$ (and thus the pitch) is determined by microscopic
intermolecular torques~\cite{Harr:1997,Luben:1997} and depends on the
molecular chirality of CLC consistuent mesogenes.  In experiments it
can be influenced either by variations in thermodynamic parameters
such as temperature or by introducing impurities.

So far we have seen that the twisting number of the ideal helical
director configuration in unbounded CLCs equals the free twisting
number $q_0$.  This is no more the case in the presence of boundary
surfaces or external fields that generally break the translational
symmetry along the twisting axis. The case of planar CLC cell bounded
by two parallel substrates exemplifies the simplest confining geometry
and is of our primary concern in this paper.

Director configurations in the CLC cell are strongly affected by the
anchoring conditions at the substrates, so that, in general, the
helical form of the director field will be distorted.
Nevertheless, when the anchoring conditions are planar and 
out-of-plane deviations of the director can be neglected, the
configurations still have the form of the helical structure which
twist wavenumber, $q$, differs from $q_0$. 
Dependence of the twist wavenumber $q$ on the free twisting number 
$q_0$ has been previously studied in
Refs.~\cite{Pink:eng:1991,Pink:eng:1992,Pink:1992} and was found to be
discontinuous. In Refs.~\cite{Zink:1995,Zink:1999} this jump-like
behaviour was shown to manifest itself in abrupt changes of selective
transmission spectra with temperature. Different mechanisms 
behind temperature
variations of the pitch in CLC cells have been discussed in
recent papers~\cite{Bel:eng:2000,Palto:eng:2002}.

Theoretical results of 
Refs.~\cite{Pink:eng:1991,Pink:eng:1992,Pink:1992,Zink:1995,Zink:1999,Bel:eng:2000,Palto:eng:2002}
are related to the CLC cells placed between two identical substrates and
the corresponding anchoring strengths were taken to be equal. 
In this paper we consider CLC cells with different anchoring energies 
at the confining surfaces. In particular, we find that sufficiently
large asymmetry in anchoring strengths will render the dependence of
the twist wavenumber on $q_0$ continuous.

The paper is organized as follows.

Director configurations within CLC cells are analyzed in
Sec.~\ref{sec:orient-struct}. Using the one-constant approximation we
derive exact solutions of the Euler-Lagrange equations. These
solutions are expressed in terms of the Jacobian elliptic functions and
correspond to inhomogeneous conical orietational structures.
The case of the planar anchoring conditions is considered   
in Sec.~\ref{sec:pitch-vs-twist} where we present both analytical and
numerical results about 
the dependence of the twist wavenumber on the free twisting number.
Concluding remarks are given in Sec.~\ref{sec:concl}. 

\section{Orientational structures in CLC cell}
\label{sec:orient-struct}

We consider director configurations in a CLC cell 
of thickness $2l$ sandwiched
between two parallel plane substrates located at $z=-l$ and $z=l$,
so that the $z$-axis is normal to the confining surfaces.
We shall also assume that the symmetry with respect to in-plane
translations is unbroken and
the director field describing orientational
structures does not depend on the coordinates $x$ and $y$.
Its parameterization in terms of the angles $\phi$ and $\theta$
is taken in the form:
\begin{equation}
  \label{eq:dir-param}
  \uvc{n}(z)=(\cos\phi(z)\cos\theta(z),\sin\phi(z)\cos\theta(z),
\sin\theta(z))\,.
\end{equation}

Substituting the parameterization~\eqref{eq:dir-param}
into Eq.~\eqref{eq:frank} and using the one-constant approximation,
$K_i=K$,
we have the following expression for 
the bulk free energy per unit area: 
\begin{align}
  F_b=\frac{K}{2}\,\int_{-l}^{l}
\bigl[
(\theta')^2+(\phi'\cos^2\theta-q_0)^2
+(\phi'\sin\theta\cos\theta)^2
\bigr]\,\dd z,
  \label{eq:f_b-gen}
\end{align}
where the prime stands for derivative with respect to $z$.

In order to specify the anchoring conditions at the boundary surfaces
$z=\pm l$ in terms of the anchoring strengths
$W_{\pm}$ and the vectors of easy orientation
$\uvc{e}_{\pm}$ we shall write the 
surface contribution to the free energy as an anchoring energy
taken in the form of Rapini-Papoular potential:
\begin{align}
F_S=\frac{W_{+}}{2}\,\Bigl[1-\bigl(\uvc{n}(l)\cdot\uvc{e}_{+}\bigr)^2\Bigr]
+\frac{W_{-}}{2}\,
\Bigl[1-\bigl(\uvc{n}(-l)\cdot\uvc{e}_{-}\bigr)^2\Bigr]\,.
  \label{eq:f_s-gen}
\end{align}
The phenomenological anchoring parameters $W_{\pm}$ characterize the
strength of interaction between CLC molecules and the surfaces, while
the preferred orientation of the molecules at the substrates is
defined by the easy axes $\uvc{e}_{\pm}$. 

\begin{figure*}[!tbh]
\centering
\resizebox{130mm}{!}{\includegraphics*{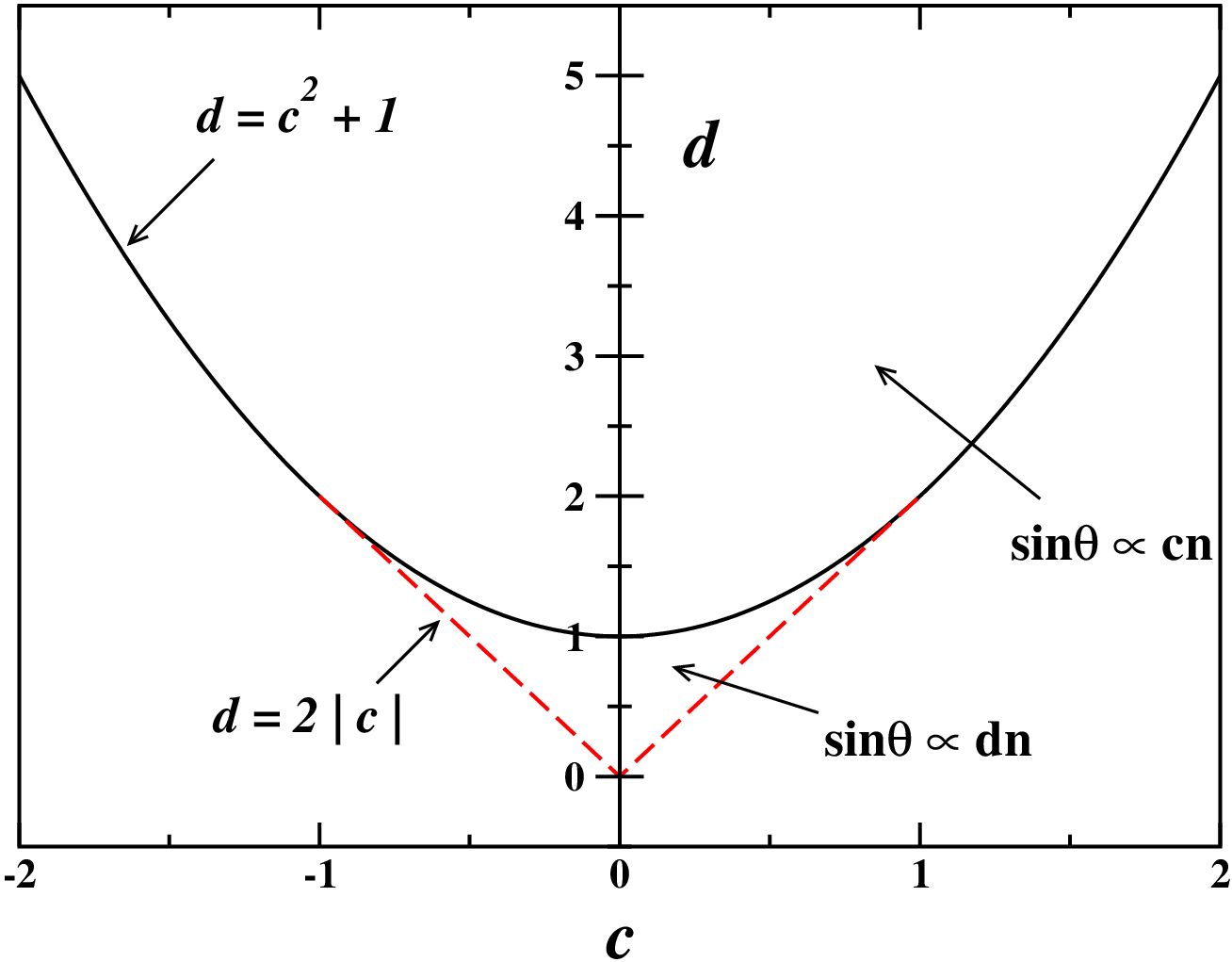}}
\caption{%
Real-valued solutions of Eq.~\eqref{eq:theta_z} exist only if
the integration constants $c$ and $d$ 
define a point in the $c$-$d$ plane that is
above the curves: $d=c^2+1$ at $|c|>1$ (solid line)  and
$d=2|c|$ at $|c|\le 1$ (dashed line).
}
\label{fig:param}
\end{figure*}

\begin{figure*}[!tbh]
\centering
\resizebox{150mm}{!}{\includegraphics*{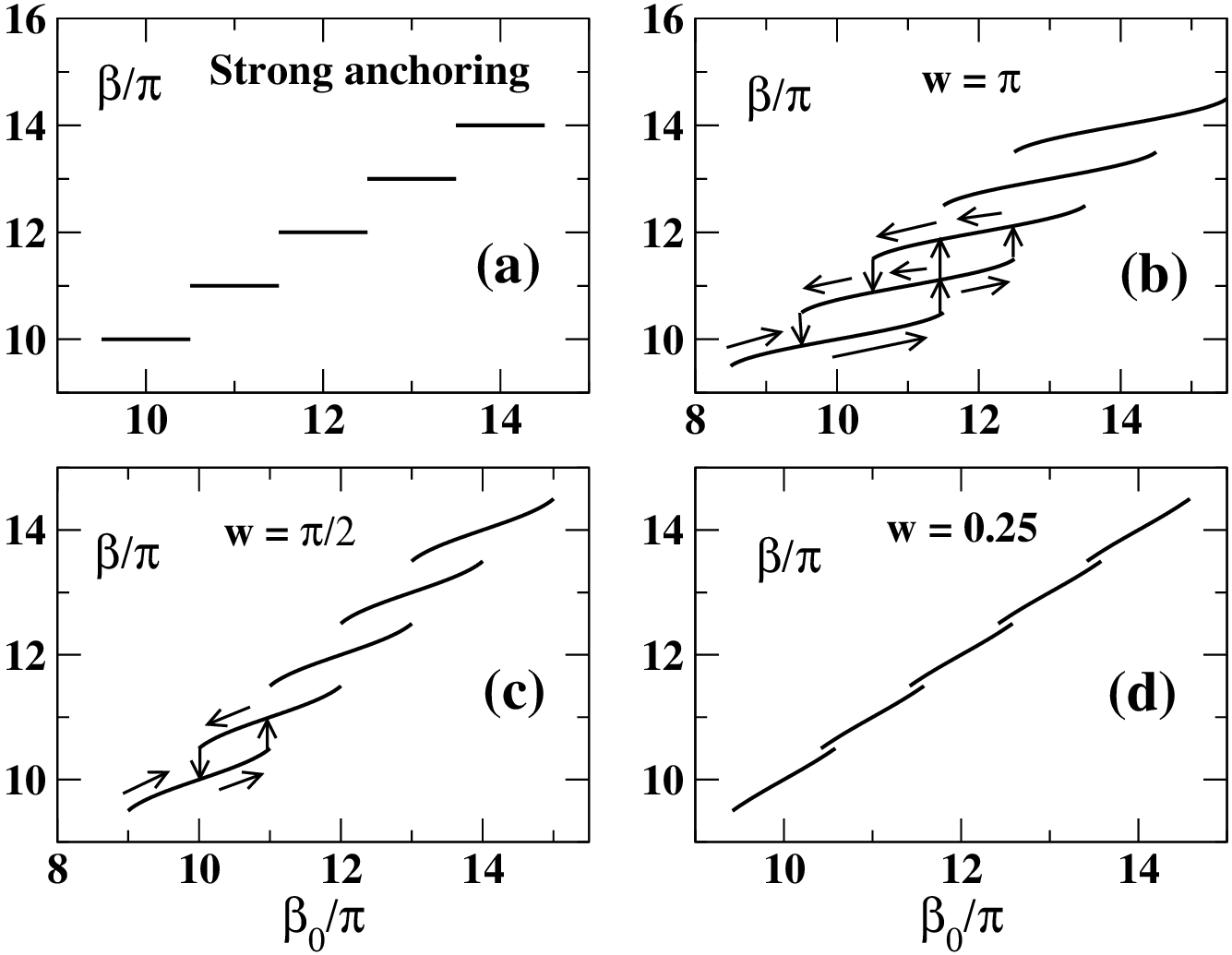}}
\caption{%
Dependencies of $\beta$ ($=2ql-\Delta\phi$) on $\beta_0$ 
($=2q_{0}l-\Delta\phi$) at $W_{-}=W_{+}\equiv W$
for various values of the dimensionless anchoring energy parameter
$w$ ($=Wl/K$): (a) strong anchoring limit, $w\to\infty$;
(b) $w=\pi$; (c) $w=\pi/2$; (d) $w=0.25$.
}
\label{fig:sym}
\end{figure*}

\begin{figure*}[!tbh]
\centering
\resizebox{150mm}{!}{\includegraphics*{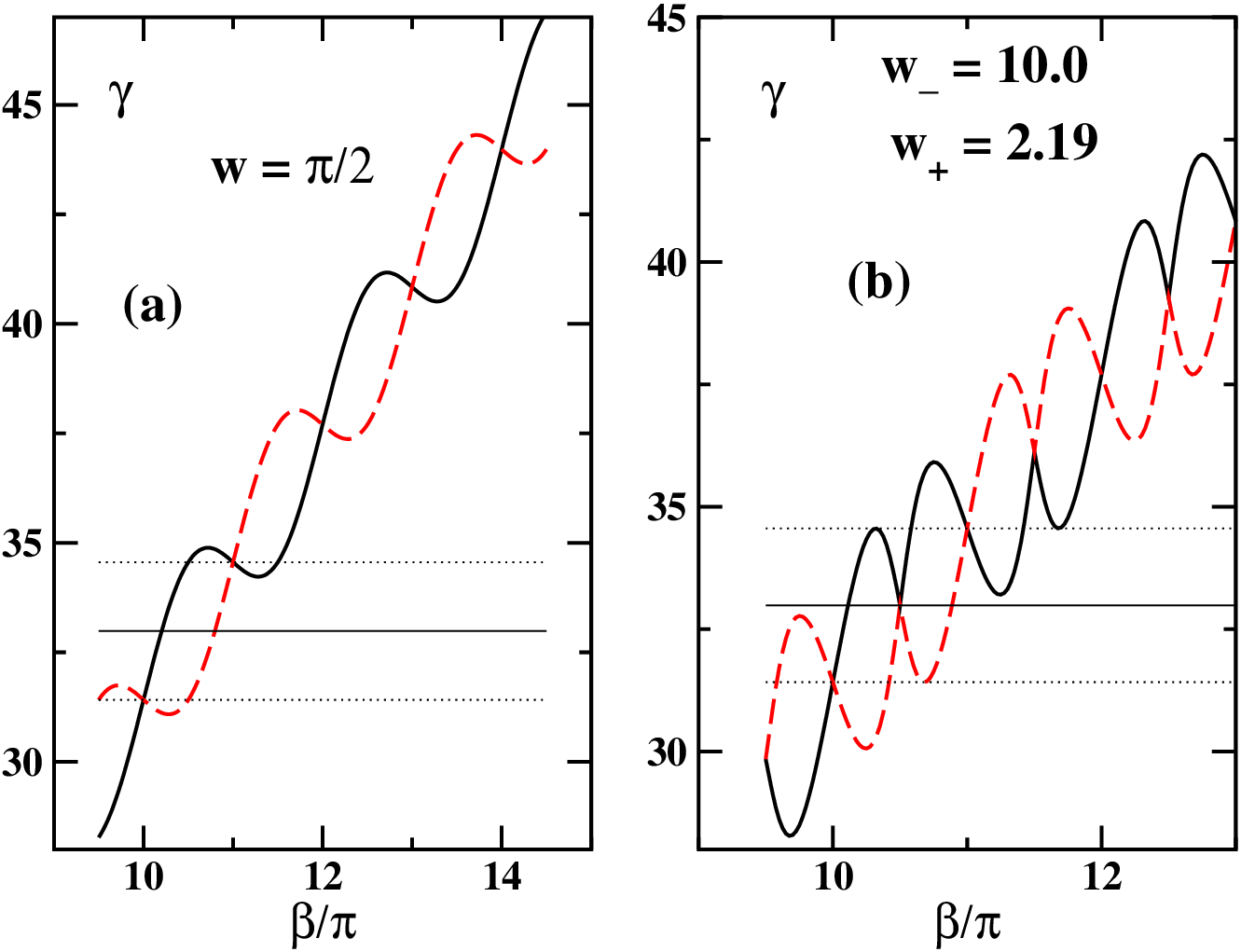}}
\caption{%
The curves
representing the plot of the function
$\gamma_{+}(\beta)$ and $\gamma_{-}(\beta)$
are shown as thick solid and dashed lines,
respectively.
The points located at the intersection of 
the curves and the horizontal straight line
$\gamma=\beta_0$ give the roots of Eq.~\eqref{eq:beta}.
The value of $\beta_0$ is $(10+1/2)\pi$
(thin solid line) and is $(10+1/2)\pi\pm w$
(thin dotted lines). 
Two cases are illustrated:
(a) $w_{+}=w_{-}=w=\pi/2$; 
(b) $w_{-}=10.0$, $w_{+}=2.19$, 
$w=\pi/2$ (see Eq.~\eqref{eq:ww}). 
}
\label{fig:inter}
\end{figure*}

\begin{figure*}[!tbh]
\centering
\resizebox{150mm}{!}{\includegraphics*{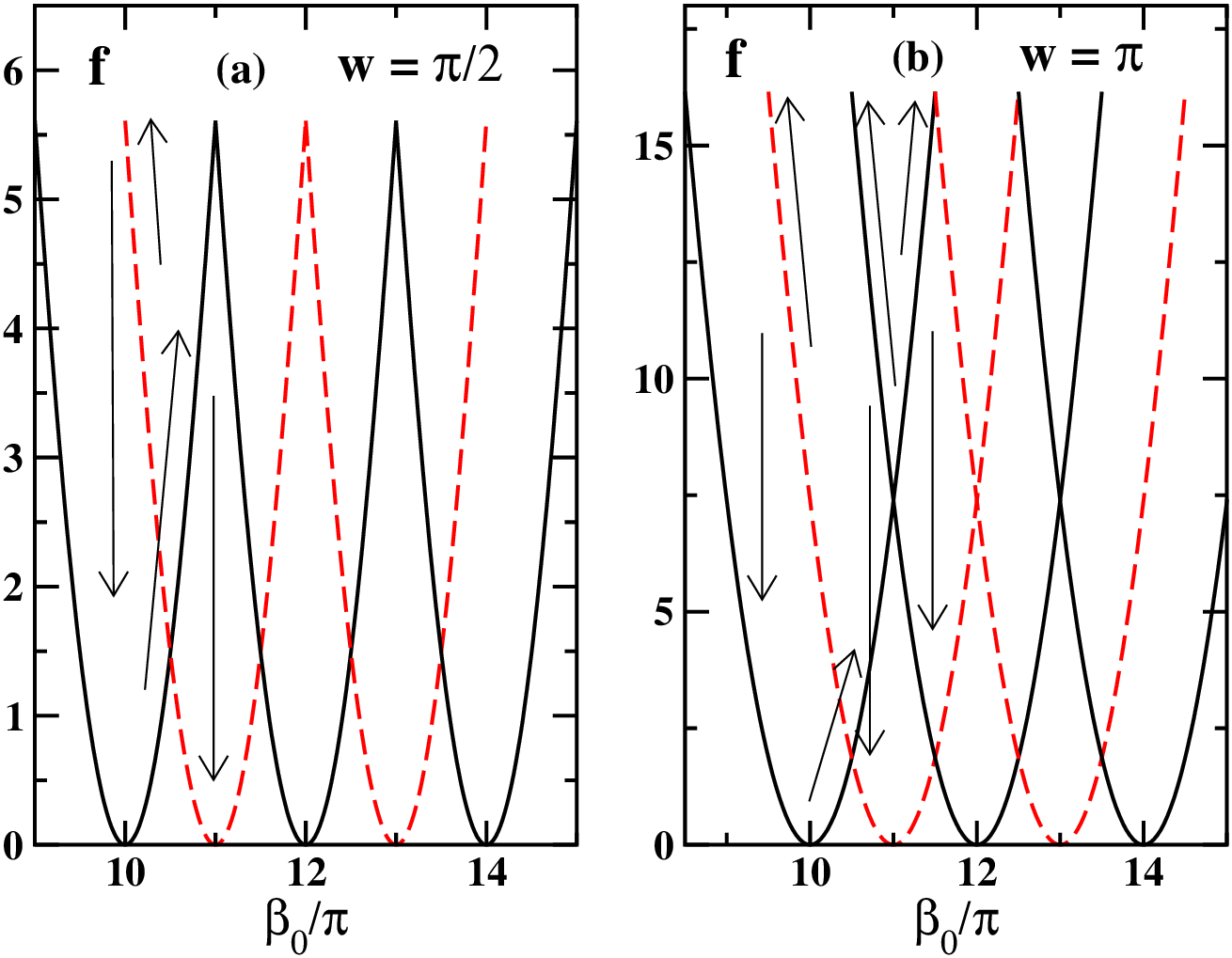}}
\caption{%
The free energy $f_{+}$ (solid line) and $f_{-}$ (dashed line)
of stable configurations with the half-turn numbers
between $10$ and $14$
as a function of $\beta_0$ computed from Eq.~\eqref{eq:f-ener-pl}
by using Eq.~\eqref{eq:beta} 
for various values of the anchoring energy parameters
(a) $w=\pi/2$;
(b) $w=\pi$.
}
\label{fig:en-bet0}
\end{figure*}

\begin{figure*}[!tbh]
\centering
\resizebox{150mm}{!}{\includegraphics*{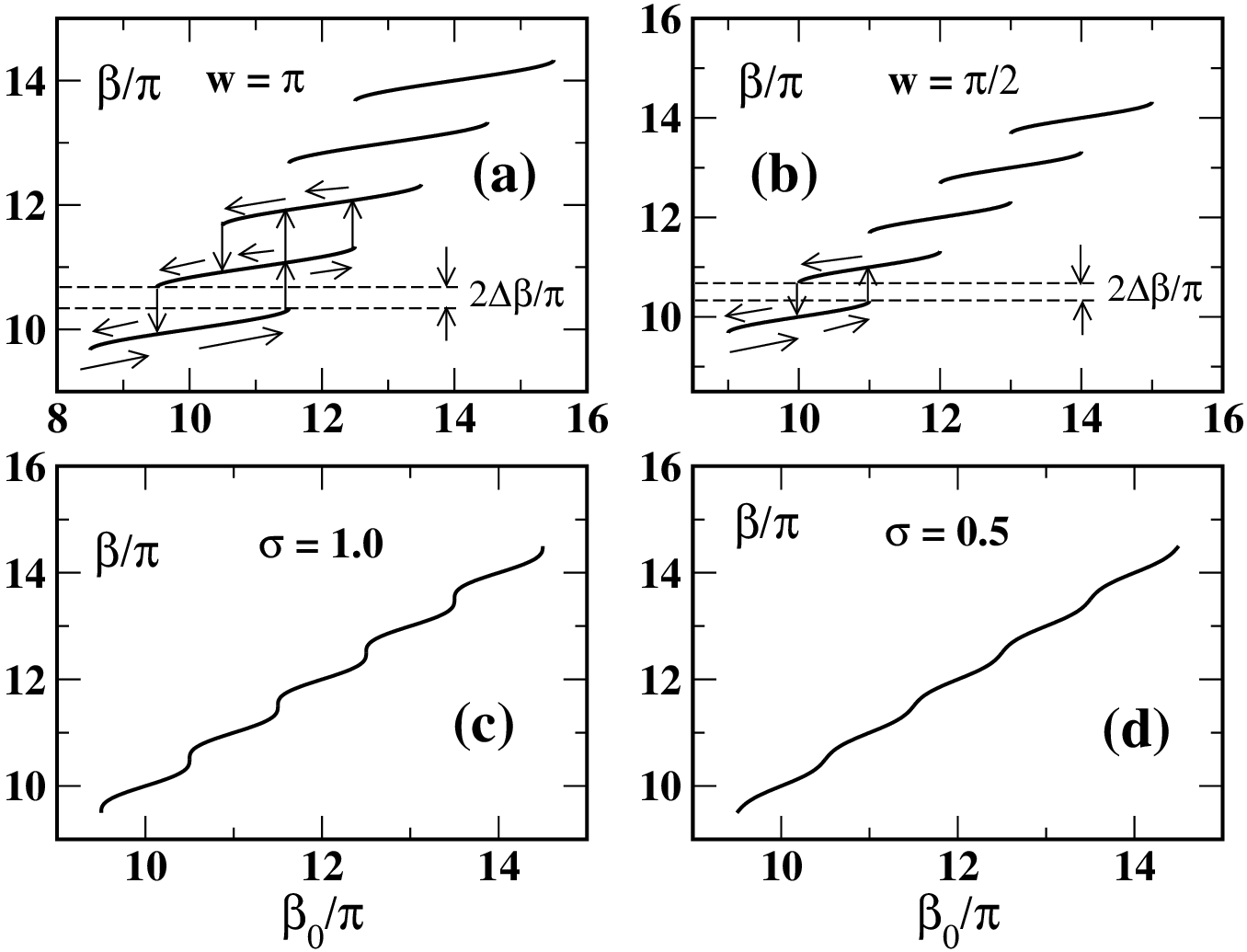}}
\caption{%
Dependencies of $\beta$  on $\beta_0$ 
calculated at $w_{-}=10.0$
for various values of the parameters
$\sigma$ ($=2w_{-}w_{+}/(w_{-}-w_{+})$) and $w$ (see Eq.~\eqref{eq:ww}):
(a) $w=\pi$ ($\sigma= 11.8$ and $w_{+}= 3.705$);
(b) $w=\pi/2$ ($\sigma= 5.6$ and $w_{+}= 2.19$);
(c) $\sigma=1.0$ ($w_{+}=w_c= 0.476$);
(d) $\sigma=0.5$ ($w_{+}= 0.25$).
}
\label{fig:asym}
\end{figure*}

\begin{figure*}[!tbh]
\centering
\resizebox{150mm}{!}{\includegraphics*{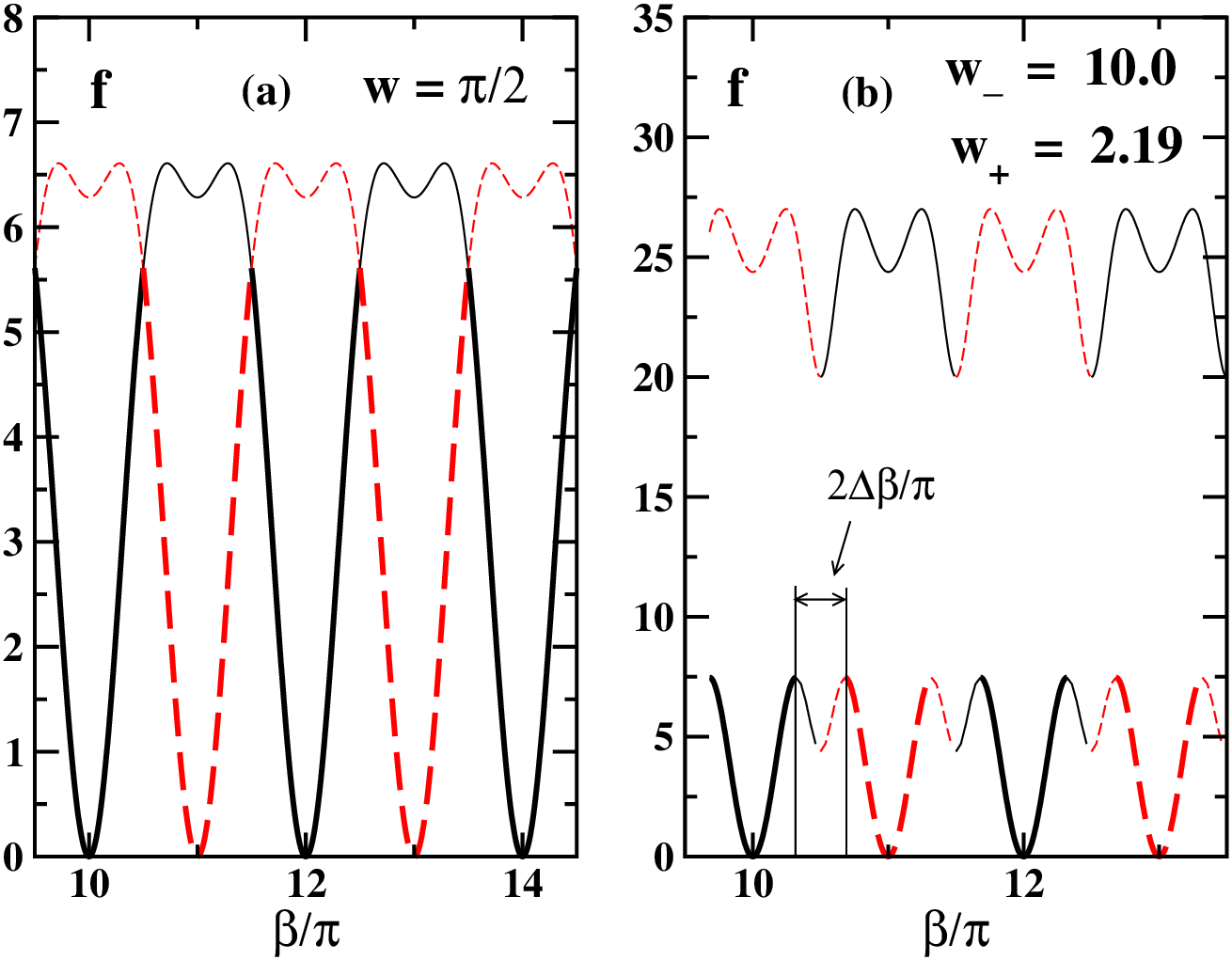}}
\caption{%
The free energy $f_{+}$ (solid line) and $f_{-}$ (dashed line)
of the configurations with the half-turn numbers
between $10$ and $14$
as a function of $\beta$ computed from Eq.~\eqref{eq:f-ener-pl}.
Thin lines represent the energy of unstable configurations.
Two cases are shown: 
(a) $w_{-}=w_{+}=w=\pi/2$;
(b) $w_{-}=10.0$, $w_{+}=2.19$
 ($\sigma=5.6$ and $w=\pi/2$).
} 
\label{fig:en-bet}
\end{figure*}

\begin{figure*}[!tbh]
\centering
\resizebox{150mm}{!}{\includegraphics*{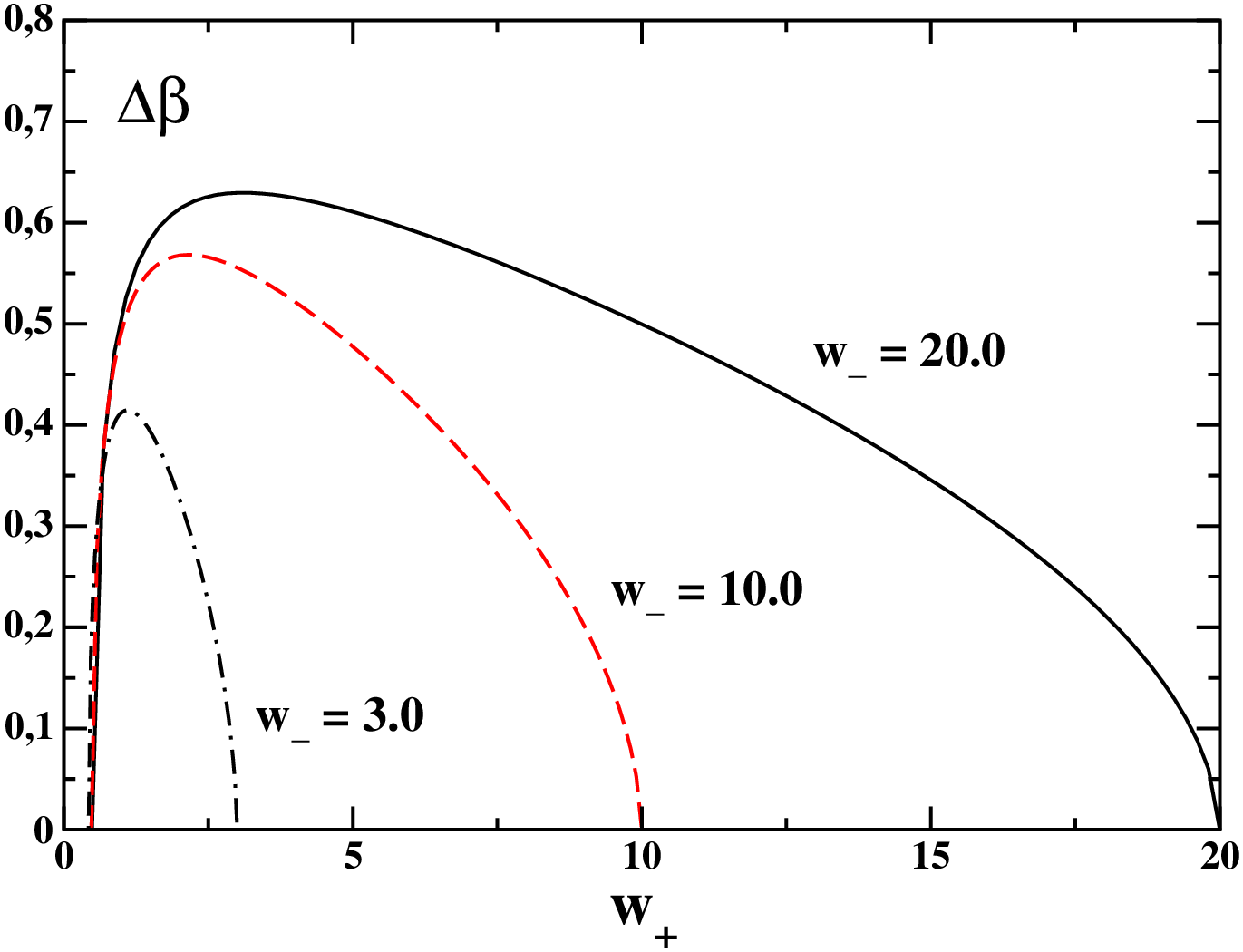}}
\caption{%
Dependencies of $\Delta\beta$  on $w_{+}$ 
for various values of the anchoring energy parameter
$w_{-}$.
}
\label{fig:delt-w}
\end{figure*}

\begin{figure*}[!tbh]
\centering
\resizebox{150mm}{!}{\includegraphics*{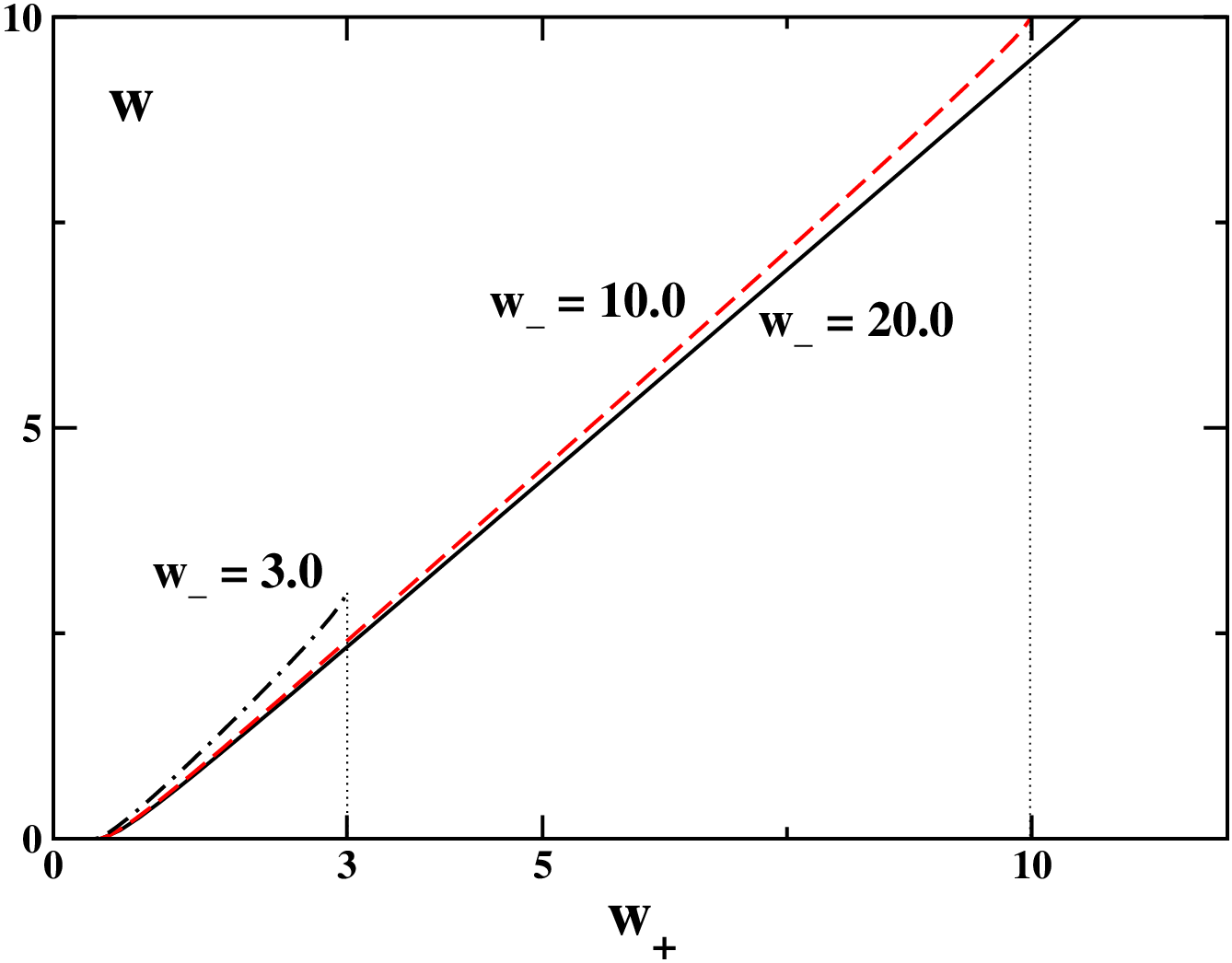}}
\caption{%
Dependencies of $w$  on $w_{+}$ 
for various values of the anchoring energy parameter
$w_{-}$. It is shown that $w=w_{-}$ at $w_{+}=w_{-}$
and $w=0$ at $w_{+}=w_c$ (see Eq.~\eqref{eq:w_crit}).
}
\label{fig:ww-w}
\end{figure*}

\subsection{Solutions of the Euler-Lagrange equations}
\label{subsec:euler-lagr-equat}

The free energy functional~\eqref{eq:f_b-gen} only depends on the
derivative $\phi'$, so that the corresponding Euler-Lagrange equation
can be integrated to yield the following equation:
\begin{equation}
\phi'=q_0 (1+c\,\cos^{-2}\theta)\,,
  \label{eq:phi_z}
\end{equation}
where $c$ is the integration constant.
Variation of the functional~\eqref{eq:f_b-gen}
with respect to $\theta$ will give the second Euler-Lagrange equation
that can be simplified by using Eq.~\eqref{eq:phi_z}. The result is as
follows
\begin{equation}
  \theta''=-q_0^2\sin\theta\cos\theta (c^2\cos^{-4}\theta-1)\,.
\label{eq:theta_zz}  
\end{equation}
The first integral of
Eq.~\eqref{eq:theta_zz} is not difficult to find and is given by
\begin{equation}
(\theta')^2+q_0^2 (c^2 \cos^{-2}\theta+\cos^2\theta)=q_0^2 d\,,
  \label{eq:1st-int}  
\end{equation}
where $d$ is the integration constant. 
From Eqs.~\eqref{eq:phi_z} and~\eqref{eq:theta_zz}
we have the bulk free energy in the form
\begin{equation}
2F_b/K=q_0^2 \int_{-l}^{l} (d-2\cos^2\theta+1) \dd z\,.
  \label{eq:fr-en-sol}
\end{equation}

We can now proceed with solving Eq.~\eqref{eq:1st-int}
that can be conveniently rewritten in the form:
\begin{equation}
[(\sin\theta)']^2=-q_0^2 (\cos^2\theta-c_{-})(\cos^2\theta-c_{+})\,,
  \label{eq:theta_z}
\end{equation}
where $2 c_{\pm}=d\pm\sqrt{d^2-4 c^2}$. This equation will have
solutions only if the values of $c_{-}$ are non-negative and 
do not exceed the unity, $0\le c_{-}\le 1$. This immediately places
the restriction on the parameter $d$: $d\ge\max(2|c|, c^2+1)$.
Fig.~\ref{fig:param} shows this region in the $c$-$d$ parameter plane.

In the upper part of the region, $d>c^2+1$, we have $c_{+}> 1$
and solutions of Eq.~\eqref{eq:theta_z} can be expressed
in terms of the Jacobian elliptic functions~\cite{Abr} as follows
\begin{equation}
  \label{eq:sol-cn}
  \sin\theta= a\cn(b q_0 z + \theta_0|m^{-1})\,,
\end{equation}
where $a^2=1-c_{-}$, $b^2=c_{+}-c_{-}$, $m=b^2/a^2$ and
$\theta_0$ is the integration constant.
Below the parabola $d=c^2+1$, where
$2|c|<d<c^2+1$ and $c_{+}<1$, these solutions are given by
\begin{equation}
  \label{eq:sol-dn}
  \sin\theta= a\dn(a q_0 z + \theta_0|m)\,.
\end{equation}
In addition, there are homogeneous solutions
that are independent of $z$: $\sin\theta=\pm\sqrt{1-c_{-}}$
and $\sin\theta=\pm\sqrt{1-c_{+}}$ (at $c_{+}<1$).
These solutions correspond to a conical helix in which the director
makes an oblique angle with respect to the helix axis.
There are no spatially varying solutions at the boundary of
the region depicted in Fig.~\ref{fig:param}. In this case the pretilt
angle does not depend on $z$ and we have
\begin{equation}
\sin\theta=
\begin{cases}
0, & b=c^2+1 \text{ and } |c|\ge 1,\\
\pm\sqrt{1-c^2}, & b=2|c| \text{ and } |c|< 1.
\end{cases}  
\label{eq:sol-bound}  
\end{equation}

Given the pretilt angle as a function of $z$, 
integrating of Eq.~\eqref{eq:phi_z} will give
the azimuth angle $\phi$:
\begin{equation}
\phi(z)=q_0 z+ c
\int_{0}^{z}\frac{q_0\dd z'}{1-\sin^2\theta(z')}+\phi_0\,,
  \label{eq:phi-sol}  
\end{equation}
where $\phi_0$ is the twisting angle in the middle of CLC cell.

In order to compute the integration constants
$c$, $d$, $\theta_0$ and $\phi_0$
we need to know the boundary conditions. The standard variational
procedure will yield the following result:
\begin{align}
&
K q_0 c =\pm W_{\pm}\cos\theta 
(\uvc{n}\cdot\uvc{e}_{\pm})
(\uvc{n}_{\phi}\cdot\uvc{e}_{\pm})|_{z=\pm l}\,,
  \label{eq:bc-phi-gen}
\\
&
K \theta'|_{z=\pm l} =\pm W_{\pm}
(\uvc{n}\cdot\uvc{e}_{\pm})
(\uvc{n}_{\theta}\cdot\uvc{e}_{\pm})|_{z=\pm l}\,,
\label{eq:bc-theta}
\end{align}
where 
$\uvc{n}_{\phi}=(-\sin\phi,\cos\phi,0)$
and $\uvc{n}_{\theta}=(-\sin\theta\cos\phi,-\sin\theta\sin\phi,\cos\theta)$.

\section{Pitch wavenumber versus free twisting number}
\label{sec:pitch-vs-twist}

In this section we consider  the simplest case of an ordinary
spiral director configuration in which the pretilt angle equals zero,
$\sin\theta=0$ and $d=c^2+1$. 
This case occurs when the anchoring conditions are
planar and both of the easy axes are parallel to the
confining surfaces. 
We direct the $x$-axis along the vector $\uvc{e}_{-}$, while
the easy axis at the upper substrate is rotated through the angle
$\Delta\phi$: $\uvc{e}_{+}=(\cos\Delta\phi,\sin\Delta\phi,0)$.

From Eq.~\eqref{eq:phi-sol} we have
\begin{equation}
\phi=q z + \phi_0\,,
  \label{eq:phi-plan}
\end{equation}
where $q=q_0(1+c)$ is the twisting wavenumber. The boundary
conditions~\eqref{eq:bc-phi-gen} can be conveniently rewritten
in the form:
\begin{align}
&
\beta-\beta_{0} = - w_{\pm}\sin(\beta\pm\alpha)\,,
\quad w_{\pm}\equiv W_{\pm} l/K\,,
  \label{eq:bc-plan}
\\
&
\beta= 2 q l -\Delta\phi,\quad
\alpha= 2\phi_0 -\Delta\phi,\quad
\beta_0= 2 q_0 l -\Delta\phi\,.
  \label{eq:bc-plan-aux}
\end{align}
Eqs.~\eqref{eq:bc-plan} define stationary points
of the free energy written as a function of the parameters
$\alpha$ and $\beta$:
\begin{equation}
    4 l F(\alpha,\beta)/K=
(\beta-\beta_0)^2 - 
w_{+}\cos(\beta+\alpha)-w_{-}\cos(\beta-\alpha)+
(w_{-}+w_{+}).
  \label{eq:f-ener-surf}
\end{equation}
There are two additional conditions for
a stationary point to be a local minimum of the free energy
surface~\eqref{eq:f-ener-surf}. These conditions ensure
local stability of the corresponding director configurations and can
be written in the following form:
\begin{align}
  &
A\equiv w_{+}\cos(\beta+\alpha)+w_{-}\cos(\beta-\alpha)>0\,,
  \label{eq:cond-min1}
\\
&
H\equiv A+2 w_{+} w_{-}
\cos(\beta+\alpha)\cos(\beta-\alpha)>0 .
  \label{eq:cond-min2}
\end{align}

From Eq.~\eqref{eq:bc-plan} we can now relate
the angles $\alpha$ and $\beta$ through the following equation
\begin{equation}
\alpha = \arctan[
\epsilon \tan\beta
] + \pi k\,,\quad
\epsilon=\frac{w_{-}-w_{+}}{w_{-}+w_{+}}\,,
  \label{eq:al-bet}
\end{equation}
where the number $k$ is integer, $k\in\mathbb{Z}$,
that is the number of half-turns of the spiral. 
Substituting the relation~\eqref{eq:al-bet} into Eq.~\eqref{eq:bc-plan}
will give a pair of transcendetial equations for $\beta$:
\begin{equation}
\beta_0=\gamma_{\pm}(\beta)\equiv\beta\pm
w_{+} \sin\bigl(
\beta +
\arctan[
\epsilon \tan\beta
] 
\bigr)\,.
  \label{eq:beta}
\end{equation}
The right hand side of Eq.~\eqref{eq:beta} is $\gamma_{+}(\beta)$
($\gamma_{-}(\beta)$) when the number $k$ is even (odd). 
Solving Eqs.~\eqref{eq:beta} will provide, in general, various values
of $\beta$. 
The values 
that meet the stability conditions~\eqref{eq:cond-min1}
and~\eqref{eq:cond-min2} correspond to metastable configurations of
CLC.  The equilibrium director structure is
determined by the solutions of Eq.~\eqref{eq:beta} 
that give the least value of the free energy.
From Eqs.~\eqref{eq:f-ener-surf} and~\eqref{eq:beta}
the free energy can be expressed in terms of the angle $\beta$
and is given by
\begin{align}
&
  4 l F_{\pm}(\beta)/K\equiv f_{\pm}(\beta) =
w_{+}^2 \sin^2 \psi_{+}\mp
(w_{+}\cos\psi_{+}+w_{-}\cos\psi_{-})+
(w_{-}+w_{+}),
  \label{eq:f-ener-pl}
\\
&
\psi_{\pm}\equiv
\beta \pm
\arctan[
\epsilon \tan\beta
]\,.
  \label{eq:psi}
\end{align}
Similarly, we can use Eq.~\eqref{eq:al-bet} to rewrite the stability
conditions as follows
\begin{align}
  &
A_{\pm}\equiv \pm (w_{+}\cos\psi_{+}+w_{-}\cos\psi_{-}) >0\,,
  \label{eq:cond-st1}
\\
&
H_{\pm}\equiv A_{\pm}+2 w_{+} w_{-}
\cos\psi_{+} \cos\psi_{-} > 0 .
  \label{eq:cond-st2}
\end{align}

It is now our task to study the dependence of 
the pitch wavenumber $q$
on the free twist wavenumber $q_0$, which is proportional
the chiral parameter, $q_0=h/K$, at different anchoring conditions.
Equivalently, we concentrate on the dependence of $\beta$
on $\beta_0$ (see Eqs.~\eqref{eq:bc-plan-aux} and~\eqref{eq:beta}).
This dependence can be thought of as a sort
of dispersion relation.

\subsection{Strong anchoring limit}
\label{subsec:strong-anch}

First we consider the case of the strong anchoring limit,
$W_{\pm}=W\to\infty$, that can be readily treated without recourse to
numerical analysis.  In this case the boundary conditions require the
director at the substrates to be parallel to the corresponding easy
axes, $\uvc{n}(\pm l)\parallel \uvc{e}_{\pm}$.  This imposes the
restriction on the values of $q$, so that $q$ takes the values from
a discrete set.  This set represents locally stable director
configurations labelled by the half-turn number $m$ as follows
\begin{equation}
2ql=\Delta\phi+\pi m,\quad m\in\mathbb{Z}\,.
  \label{eq:ql-str}
\end{equation}
Substituting the values of
$\beta$ from the relation~\eqref{eq:ql-str} into the first term
on the right hand side of Eq.~\eqref{eq:f-ener-pl} will define
the equilibrium value of $m$ as the integer  that minimizes the distance
between $\pi m$ and $2 q_0 l - \Delta\phi$.
The resulting step-like dependence is depicted in Fig.~\ref{fig:sym}a.
 
\subsection{Equal anchoring energies}
\label{subsec:equal-anch}

When the anchoring strengths at both  substrates are equal,
$W_{-}=W_{+}\equiv W$, the right hand side of Eq.~\eqref{eq:beta}
is  $\beta\pm w\sin\beta$ ($w=Wl/K$)
and $\psi_{+}=\psi_{-}=\beta$. So, as is illustrated in
Fig.~\ref{fig:inter}(a), we need to find the intersection points
of the horizontal line $\gamma=\beta_0$ 
and the curves $\gamma=\gamma_{\pm}(\beta)$ in the $\beta-\gamma$
plane. The stability conditions $A_{+}>0$ and $H_{+}>0$ 
($A_{-}>0$ and $H_{-}>0$) require the values of  
$\beta$ to be ranged  between
$(m-1/2)\pi$ and $(m+1/2)\pi$, where $m$ is even (odd) integer
equal to the half-turn number.
The function $\gamma_{+}$ ($\gamma_{-}$) monotonically increases 
on these intervals, so that the value of $\beta_0$ runs from
$(m-1/2)\pi-w$ to $(m+1/2)\pi+w$ on the interval
with the half-turn number $m$: $[(m-1/2)\pi,(m+1/2)\pi]$.
For each number of $m$ we have the monotonically increasing
branch of the $\beta$ vs $\beta_0$ dependence.
The branches with $m$ ranged from $10$ to $14$ for
different values of the dimensionless anchoring energy parameter
$w$ are depicted in Fig.~\ref{fig:sym}. 
It is illustrated that the $\beta_0$-dependence of $\beta$ will be
discontinuous provided the anchoring energy is not equal to zero.
Fig.~\ref{fig:sym} shows that the jumps tend to disappear
in the limit of weak anchoring, where the anchoring energy
approaches zero, $w\to 0$.  

Similar to the case of strong anchoring, at $\beta_0=\pi/2+\pi m$
we have two different roots of Eq.~\eqref{eq:beta}
with $k=m$ and $k=m+1$
that are equally distant from $\beta_0$ and are of equal energy.
According to Refs.~\cite{Zink:1995,Zink:1999,Bel:eng:2000,Palto:eng:2002}
it can be assumed that the jumps
actually occur at the end points $\beta_0=\pi/2+\pi k \pm w$,
where the configuration 
with $m=k+1/2\mp 1/2$ becomes marginally stable 
($ A_{\pm}=H_{\pm}=0$) and loses its stability.
In other words,
the system needs  to penetrate the barrier separating
the states with different half-turn numbers. As it is seen
from Fig.~\ref{fig:sym}, in this case the upward and backward
transitions with $\Delta m=\pm 1$:
$k\to k+1$ and $k+1\to k$ occur at different values of $\beta_0$:
$\beta_{+}^{(k)}=\pi/2+\pi k + w$ and
$\beta_{-}^{(k)}=\pi/2+\pi k - w$, correspondingly.
It means that there are hysteresis loops in 
the response of CLC to the change in the free twisting number.

When the anchoring energy is small and $w < \pi/2$,
there are only two configurations at the critical point
$\beta_0=\beta_{+}^{(k)}$: the marginally stable configuration
with $m=k$ and the equilibrium one with $m=k+1$. In this case
Eq.~\eqref{eq:beta} has at most two roots and 
the jumps will occur as transitions between the states 
which  half-turn numbers differ by the unity, $|\Delta m|=1$.

Fig.~\ref{fig:en-bet0}a illustrates that for $w=\pi/2$ we have two
marginally stable structures of equal energy with $m=k$ and $m=k+2$.
Being metastable at $\pi/2< w<\pi$ this newly formed structure and the
configuration with $m=k+1$ will be degenerate in energy at $w=\pi$. So, as
is shown in Figs.~\ref{fig:sym}b and~\ref{fig:en-bet0}b, both
transitions $k\to k+1$ and $k\to k+2$ are equiprobable,
so that we have the bistability  effect at the critical point
under $w=\pi$. 
For $\pi< w <2\pi$ 
the configuration with $m=k+2$ gives the equilibrium director
structure at $\beta_0=\beta_{+}^{(k)}$.  

It is not difficult to see
that further increase of the anchoring energy involves the
configuration with $m=k\pm l$ into the transition when the parameter
$w$ is passing through the value $(l-3/2)\pi$. The jump to the
equilibrium state will require the half-turn number changed from $k$
to $k+\Delta m$, where $|\Delta m|=l$ at $(l-1)\pi< w < l\pi$.  There
are a number of transient metastable configurations involved in
transitions with $|\Delta m|<l$.

\subsection{Different anchoring energies}
\label{subsec:diff-anch}

When the anchoring energies at the surfaces are different,
$W_{-}\ne W_{+}$, 
$\sin\psi_{+}$ on the right hand side of Eq.~\eqref{eq:beta}
equals zero at $\beta=\pi/2+\pi k$ 
and, as is seen from Fig.~\ref{fig:inter}b, we have additional
intersection points of the curves $\gamma_{+}(\beta)$
and $\gamma_{-}(\beta)$. The stationary points, where the derivative
of $\gamma_{\pm}$ with respect to $\beta$ equals zero, represent
the local maxima and minima of $\gamma_{\pm}$ and are located at 
$\beta=\pi/2+\pi k\pm \Delta\beta$. The value of $\Delta\beta$
can be calculated by solving the following equation
\begin{equation}
  \label{eq:delt-bet}
  w_{+}(1+\epsilon) \sin\bigl(\Delta\beta
-\epsilon
\arctan[
\epsilon \cot\Delta\beta]
\bigr)=
-\frac{1+(\epsilon^2-1)\cos^2\Delta\beta}{1+(\epsilon-1)\cos^2\Delta\beta}\,,
\end{equation}
where $\Delta\beta\in [0,\pi/2]$.

From the stability conditions $H_{+}>0$ 
($H_{-}>0$) the values of  
$\beta$ for stable configurations fall  between
the stationary points
$(m-1/2)\pi+\Delta\beta$ and $(m+1/2)\pi-\Delta\beta$, 
where the half-turn number $m$ is even (odd) integer.
The function $\gamma_{+}$ ($\gamma_{-}$) monotonically increases 
and $\beta_0$ varies from
$(m-1/2)\pi-w$ to $(m+1/2)\pi+w$ on these intervals.
The dimensionless parameter $w$, 
as opposed to the case of equal anchoring energies,
is now given by
\begin{equation}
  \label{eq:ww}
  w=w_{+}\cos\bigl(\Delta\beta
-\epsilon
\arctan[
\epsilon \cot\Delta\beta]
\bigr)-\Delta\beta\,.
\end{equation}

Clearly, we can now follow the line of reasoning
presented in the previous section to find out
the results concerning hysteresis loops and bistability
effects that are similar to the case of equal anchoring strengths.
There are, however, two important differences related
to Eqs.~\eqref{eq:delt-bet} and~\eqref{eq:ww}.

If $\Delta\beta\ne 0$,
the intervals of $\beta$ representing stable director configurations
are separated by the gap of the length $2\Delta\beta$.
The presence of this gap is illustrated in Figs.~\ref{fig:asym}a
and~\ref{fig:asym}b. Fig.~\ref{fig:en-bet}b shows the gap
between stable branches of the dependence of the free energy
on $\beta$. The values of $\beta$ that are within the gap represent
unstable configurations (``forbidden'' states of CLC). 
The graph of the $\Delta\beta$ vs $w_{+}$
dependence is presented in Fig.~\ref{fig:delt-w} and indicates
that the gap disappears in the limit of equal energies, $w_{+}=w_{-}$.
In addition, referring to Fig.~\ref{fig:delt-w}, there is a critical
value of $w_{+}$ below which $\Delta\beta$ equals zero.

It can be shown that
Eq.~\eqref{eq:beta} has the only root, 
$\beta=\beta_0$ at $\beta_0=\pi/2+\pi k$, under the
anchoring energies meet the following condition:
\begin{equation}
\sigma\equiv\frac{2w_{-}w_{+}}{|w_{-}-w_{+}|}\le 1\,.
  \label{eq:crit}
\end{equation}
In this case the gap disappears and the dependence of
$\beta$ on $\beta_0$ becomes continuous in the manner
indicated in Figs.~\ref{fig:asym}c and~\ref{fig:asym}d.
Given the value of $w_{-}$ the relation~\eqref{eq:crit}
yields the threshold value for 
the anchoring strength at the upper substrate.
The critical anchoring energy parameter $w_c$ is given by

\begin{equation}
w_c=\frac{w_{-}}{2 w_{-}+1}\,.
  \label{eq:w_crit}
\end{equation}

As illustrated in Fig.~\ref{fig:ww-w}, the dependence of
$w$ on $w_{+}$ is approximately linear for large values
of $w_{-}$ and $w$ goes to zero at the critical point $w_{+}=w_c$.

\section{Conclusion}
\label{sec:concl}

In this paper we have studied how the pitch wavenumber
of CLC cell depends on the free twisting number for
the cases in which the anchoring strengths at the substrates
are either equal or different.
It is found that
in both cases this dependence is generally discontinuous
and is characterized by the presence of hysteresis and bistability.
But the difference in the anchoring energies
introduces the following two effects:
\renewcommand{\theenumi}{\alph{enumi}}
\renewcommand{\labelenumi}{(\theenumi)}
\begin{enumerate}
\item 
the jump-like behaviour of the twist wavenumber is suppressed
under the anchoring strength at one of the substrates is
below its critical value;

\item 
the twist wavenumber intervals of locally stable configurations
with adjacent numbers of the helix half-turns are separated by 
the gap where the structures are unstable. 
\end{enumerate}

The part of our analysis, presented in Sec.~\ref{subsec:equal-anch},
relies on the assumption that 
the transition between configurations with different
half-turn numbers occurs when the initial structure loses its
stability, so that its pitch wavenumber no more corresponds to
a local minimum of the free energy surface. The result is that
the stronger the anchoring, the larger the change of the half-turn
number (and of the twist wavenumber) needed to reach the equilibrium
state. So, whichever mechanism of relaxation is assumed, 
the metastable states certainly play an important part in the problem.   
Dynamics of the transitions is well beyond the scope of this paper
and it still remains a challenge to develop a tractable theory
that accounts for director fluctuations, hydrodynamic modes
and defect formation. 

Analytical results of Sec.~\ref{sec:orient-struct} can be used
to study the effect of pretilt angles at the confining surfaces.
This problem requires a more complicated analysis that will be
published elsewhere.


\end{document}